\documentclass[letter]{aa}  
\usepackage{graphicx}
\usepackage{natbib}
\bibpunct{(}{)}{;}{a}{}{,}

\usepackage{txfonts}

\def\h17{H1743$-$322}
\def\igr{IGR~J17464$-$3213}
\def\ergcms{erg~cm$^{-2}$~s$^{-1}$}
\def\nh{$N_{\rm H}$}
\newcommand{\rxte}{\textsl{RXTE}}
\newcommand{\inte}{\textsl{INTEGRAL}}
\newcommand{\xmm}{\textsl{XMM/Newton}}
\newcommand{\swift}{\textsl{Swift}}
\newcommand{\ariel}{\textsl{Ariel V}}
\newcommand{\heao}{\textsl{HEAO I}}

\begin{document}

   \title{The early phase of a \h17\ outburst observed by \inte, \rxte, \swift, and \xmm}
   \authorrunning{L. Prat et al.}
   \titlerunning{An early-phase hard state outburst of \h17}

   \author{L. Prat
          \inst{1,2}
          \and
          J. Rodriguez\inst{1,2}
          \and
          M. Cadolle~Bel\inst{3}
          \and
          E. Kuulkers\inst{3}\thanks{on behalf of the INTEGRAL Galactic bulge monitoring team}
          \and
          M. Hanke\inst{4,5}
          \and\\
          J. Tomsick\inst{6}
          \and
          S. Corbel\inst{1,2}
          \and
          M. Coriat\inst{1,2}
          \and
          J. Wilms\inst{4,5}
          \and
          A. Goldwurm\inst{1,7}}

   \institute{DSM/IRFU/Service d'Astrophysique/CEA-Saclay, France\\
              \email{lionel.prat@cea.fr}
\and AIM-UMR 7158, Paris, France 
\and ESAC, ISOC, Villa{\~n}ueva de la Ca{\~n}ada, Madrid, Spain
\and Dr. Karl Remeis-Observatory, Bamberg, Germany
\and Erlangen Centre for Astroparticle Physics, Germany
\and Space Sciences Laboratory, University of California, Berkeley, USA
\and APC-UMR 7164, Paris, France}

   \date{Received ; accepted }

 
  \abstract
   {}
   {We investigate the early phase of the first state change during the 2008 September-November outburst of H1743-322, first detected by the \inte\ satellite. We focus on the preliminary hard X-ray state with the aim of investigating the possible influence of this phase on the subsequent evolution during the outburst.}
   {The outburst started on MJD 54732, and remains ongoing at the time of writing this paper (MJD 54770). We analyse \inte, \rxte, \swift, and \xmm\ observations, which provide coverage of the quiescence to outburst evolution in the 3--200 keV range every few days. We present both the spectral and timing analysis. We compare these parameters with the behaviour of the source during a previous outburst in 2003, which was observed by \inte\ and \rxte.}
   {The energy spectra are well fitted by a phenomenological model consisting of an exponentially cut-off power law plus a disc component. A more physical model of thermal Comptonisation (and a disc) represents the spectra equally well. In a first phase (up to MJD 54760), the photon index and temperature of the disc do not vary significantly, and have values reminiscent of the Hard State (HS). The timing analysis is also consistent with that of a HS, and shows in particular a rather high degree of variability ($\sim$30\%), and a strong $\sim$0.5--1 Hz QPO with its first harmonic. After MJD 54760, a change to softer spectra and a $\sim$5--6 Hz QPO indicate that the source underwent a state transition into a Hard-Intermediate State (HIMS).}
   {The timing and spectral characteristics of \h17\ are similar to those of the first HS during its 2003 outburst. We observe a correlation between the QPO frequency and the photon index, which indicates a strong link between the accretion disc, generally understood to determine the QPO frequency, and the corona, which determines the QPO power. The gradual disappearance of the QPO harmonic, and the slowly decreasing hard X-ray flux, imply that the accretion disc slowly moved inwards during the HS.}


   \maketitle
%

\section{Introduction}

The X-ray nova \h17 was discovered during a bright outburst that occurred in 1977 with the \ariel\ and \heao\ satellites by \citet{Kaluzienski:1977}. In 2003, another bright outburst was first detected with \inte, and the source was then dubbed \igr, before it was realised that it was \h17 \citep{Markwardt:2003}. This outburst has been studied considerably at all wavelengths \citep[see e.g.][]{Parmar:2003, Lutovinov:2005, Kalemci:2006}. It was shown in particular that \h17\ had a behaviour consistent with most black-hole X-ray transients, and was, thus, classified as a Black-Hole Candidate. This 2003 outburst was followed by weaker episodes in 2004, 2005, and in the first months of 2008 \citep[see][]{Kalemci:2008}.

A new outburst was detected on 2008 September 23 \citep{Kuulkers:2008}, during \inte\ observations of the Galactic Bulge (GB). The source brightened above the detection limit in the middle of the \inte\ observation, and thus provided us with the rare privilege of observing an X-ray nova during the very first stages of an outburst, from the quiescent state to the rising phase. Due to this early notice, several high energy satellites were able to follow rapidly \h17: \inte\ and \rxte\ observations occurred almost every second day, while at softer X-rays, \swift\ (XRT) and \xmm\ provided 3 and 1 observations, respectively.

Herein, we present the results of the X and $\gamma$--ray coverage of the source in its Hard State (HS) until its transition to the Hard Intermediate State \citep[HIMS, see e.g.][]{Homan:2005}, which is characterized by a photon index between $\sim$1.5-2.5, and the presence of strong type C Quasi Periodic Oscillations (QPOs). Our study is focused on the very first stages of its evolution from quiescence to outburst. We analyse the spectral and timing characteristics of \h17, and compare its behaviour with that of the well-studied 2003 outbust. The outburst remains ongoing while this letter is being written, and the study of the complete outburst is deferred to a future investigation.

\section{Observations and data reduction}

\begin{figure}
   \centering
   \includegraphics[width=0.4\textwidth]{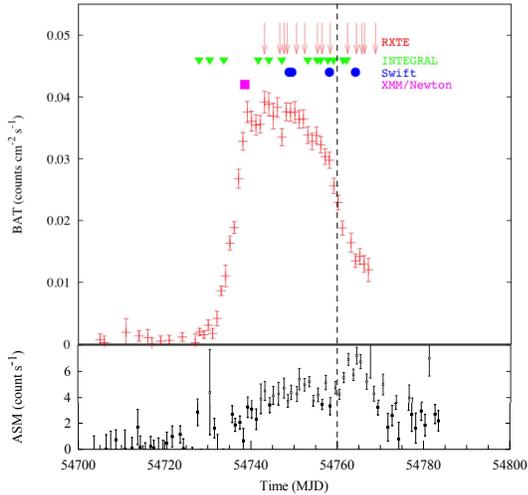}
      \caption{\swift/BAT and \rxte/ASM lightcurves of \h17\ during its 2008 outburst, in the 15-150 keV and 1.5-12 keV energy ranges, respectively. The periods of the observations used in this paper are indicated for each instrument.}
         \label{resume}
\end{figure}

The \inte\ observations analysed here originate in two different programmes. The Galactic Bulge (GB) programme \citep{Kuulkers:2007} is performed once every \inte\ revolution whenever the region is visible, and consists of 12.6 ks exposure observations. The Galactic Centre (GC) programme provides longer exposures, of duration typically between $\sim$40 and $\sim$200 ks, every $\sim$2 revolutions, as part of the ``Key Projects" strategy of \inte. The black-hole candidates within the field of view are monitored by our team \citep[PI Prat, see][]{Prat:2008}, by taking advantage of the long exposure time to achieve early detections of new outbursts and monitor their evolution. The \xmm\ observation originate in a private programme \citep[PI Wilms,][]{Hanke:2008}, while the data from \rxte\ and \swift\ are public. The times of all observations are indicated on Fig. \ref{resume}.

\medskip

\begin{figure}
   \centering
   \includegraphics[angle=-90,width=0.46\textwidth]{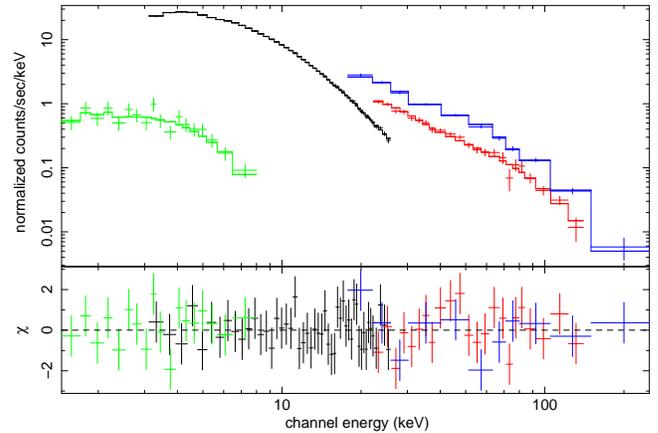}
     \caption{Joint \rxte/\inte\ spectrum, taken around MJD 54748 (\inte\ revolution 731), with the PCA (black), HEXTE (red), IBIS/ISGRI (blue) and \swift/XRT (green) data, along with the best-fit model: an absorbed powerlaw with a high-energy cutoff, a disc blackbody and a faint Fe fluorescent line. The residuals, in $\sigma$ units, are also shown below.}
         \label{exemple}
\end{figure}

\begin{figure}[!ht]
   \centering
   \includegraphics[width=0.47\textwidth]{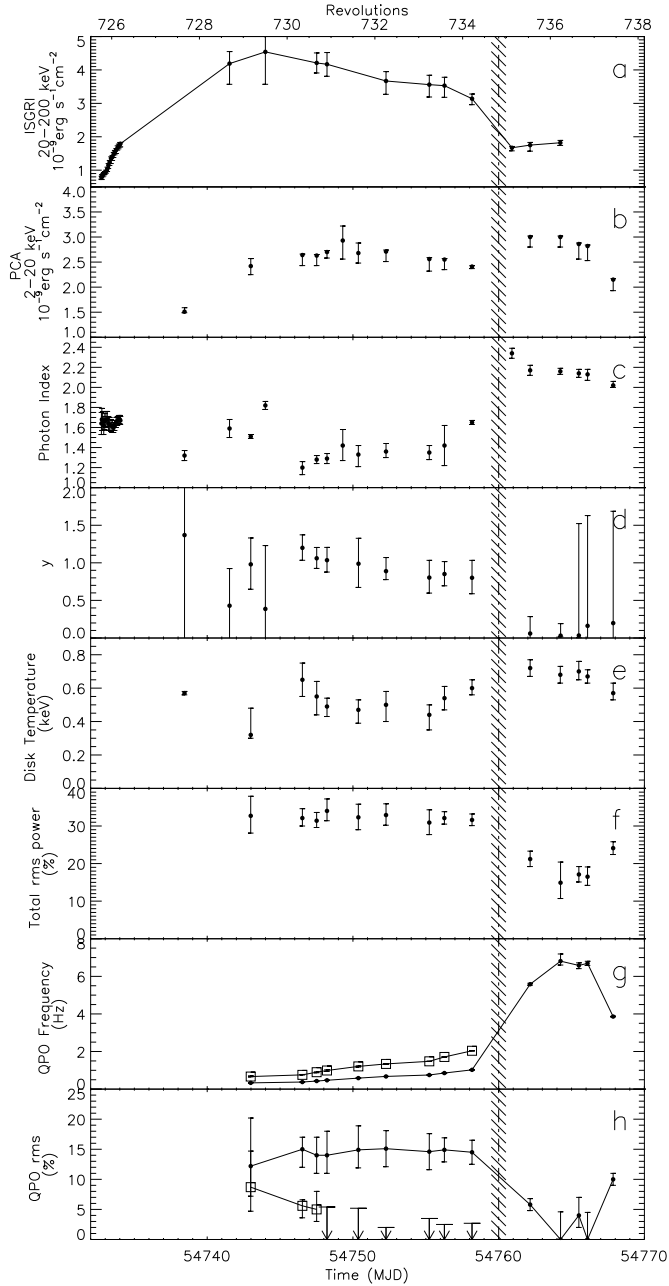}
      \caption{Spectral and timing characteristics of \h17. From top to bottom, lightcurves in the 20--200 keV (a) and 2--20 keV (b) bands, (c) photon index of the power law component, (d) Kompaneets parameter, (e) disc blackbody maximum temperature, (f) total RMS power, (g) frequencies of the two detected QPOs and (h) RMS power of these QPOs. The filled zone marks the spectral transition from the Hard State (HS) to the Hard Intermediate State (HIMS). Error bars are at the 90\% confidence level. }
         \label{results}
\end{figure}

The \inte\ data were reduced using the standard {\tt{Off-line Scientific Analysis (OSA)}} v7.0 software package \citep{Goldwurm:2003}. We used standard response files to obtain spectra of the source in the 18-200 keV range, with 11 spectral bins (see Fig. \ref{exemple} for a sample spectrum). A systematic error of 2\% was applied to all bins. The activation of \h17\ was monitored for $\sim$170 ks without interruption by \inte, and we therefore took special care in analysing the very first data available. Since \h17\ was too faint to be detected in single science window pointings, we used a ``sliding" technique: we accumulated images of $\sim$30 ks exposure time, every $\sim$6 ks. This enables a more detailed study of its evolution, especially when considering the 20--200 keV flux (Fig. \ref{results}, panel a). In the 3--20 keV range, JEM-X data were included in our analysis only after MJD 54749, when the source was sufficiently bright for spectral extraction to be possible in the data.

The \rxte\ and \swift/XRT data were reduced with the {\tt{HEASOFT}} v6.5 software package following the standard steps explained in the RXTE cookbook\footnote{available at http://heasarc.gsfc.nasa.gov/docs/xte/data\_analysis.html} and XRT users manual\footnote{http://heasarc.gsfc.nasa.gov/docs/swift/analysis/}. High time resolution light curves were extracted from the PCA (all layers, all active PCUs)  EVENT data with $\sim$500 $\mu$s resolution, after removing the artificial time marker with {\tt{sefilter}}. We restricted the energy range to $\sim$3--40 keV (Channels 5--91). We produced power-density spectra (PDS) with {\tt{powspec}} v1.0 in the frequency range 0.0156--1024 Hz. The PCA spectra were extracted in the $\sim$3--25 keV range from the top layer of PCU 2, from STANDARD 2 data, while background spectra were estimated from the PCA background model for  bright sources. Systematic errors at a level of 0.8\% were added to all channels. The HEXTE spectra were extracted from Cluster B only.

For XRT data, we produced level 2 data with the {\tt{xrtpipeline}} v0.12.0, which removed calibration source, bad pixels and grade selection. We analyse only observations completed in photon-counting mode. We extracted images, light curves, and spectra from these cleaned event files with {\tt{xselect}} v2.4. However, even during the first observation the source was bright and its light was lightly to cause saturation; we therefore extracted source and background spectra from annuli centred on the most accurately determined position of H1743$-$322. We generated exposure maps with {\tt{xrtexpomap}} v0.2.5, and ancillary response files with the tool {\tt{xrtmkarf}} v0.5.6. The resultant spectra were rebinned so as to have at least 20 counts per channel, allowing the chi-squared statistics to be evaluated using {\tt{xspec}} v11.3.2ag. The spectra were fitted between 0.5~keV and $\sim$10~keV.

The \xmm\ data were reduced with the Science Analysis Software, \texttt{xmmsas}, v.~7.1, following standard procedures. We used successively the tasks \texttt{epchain}, \texttt{rgsproc}, \texttt{evselect}, \texttt{rmfgen}, and \texttt{arfgen} to produce spectra and response matrices. We restricted the data from the EPIC-pn (timing mode) to the 1--10\,keV range. The first and second order spectra from RGS\,1 and 2 were rebinned to contain $\ge50$ counts per bin.

\section{Results}

\subsection{Spectral analysis and results}
Apart from the first stages of the outburst, when the flux increased quickly, the source did not evolve significantly during any single day. Therefore, we combined the data from the various satellites when the observations were separated by less than one day, to be able to extend the energy range and constrain the spectral parameters more accurately.

The high-energy ends of the spectra are well fitted using a simple power-law, with a high-energy cutoff located typically in the $\sim$15--25 keV range, with relative error bars of $\sim$20\%. The 1--20 keV range shows a disc blackbody component and a Gaussian Fe K emission line. As a cut-off power law is usually interpreted as a signature of thermal Comptonization, we replaced this simpler model with a more physical Comptonization model \citep[\texttt{COMPTT,}][]{titarchuk:1994}, and obtained equally good fits. The plasma temperatures that we obtained are similar to those corresponding to the cutoff energies measured using the phenomenological model, with slightly smaller error bars. Hereafter, the photon indices were thus deduced from the phenomenological model, while the other spectral parameters were calculated using the more physical {\sc constant*wabs*(comptt+diskbb+gaussian)} model (in {\tt XSPEC} terminology). The absorption is not well constrained by the \rxte/PCA observations and, when leaving this parameter free, we were unable to determine the temperature of the disc. Therefore, we assumed the value calculated using the \swift\ and \xmm\ observations, \nh$=1.8 \pm 0.2 \times 10^{22} $cm$^{-2}$, and fixed the \nh\ parameter to this value in every observation.

The spectral evolution of \h17\ is reported in Fig. \ref{results} (panels a to e). The sizes of the error bars depend on the availability of the various satellites.

During the first days of the outburst, around MJD 57433, the source exhibited a rapid rise in hard X-ray flux: the flux increased by a factor $\sim$2.3 in less than 1.5 days (Fig. \ref{results}, panel a). At the same time, the photon index remained remarkably stable, around 1.65 (panel c). Then, between MJD 54741.5 and MJD 54760, the 20--200 keV flux reached a maximum and slowly decayed, while the photon index stayed at low values, in the range 1.2--1.6. The times when \rxte\ or \swift\ satellites are available correspond to indices lower by $\sim$0.3 compared with the $\inte$ data alone. This is due to the fact that the two former satellites provide coverage in softer energy bands: since the coverage of ISGRI starts around 18 keV, it cannot accurately constrain the high-energy cutoff around $\sim$20 keV. Instead, the presence of this cutoff appears to cause the photon index to be softer, because the fitting process tends to mix the beginning of the high energy rollover with the end of the pure power-law part.

The Comptonized component can be characterised by the Kompaneets parameter \citep{Rybicki:1986}, $y= \frac{4kT_e}{m_ec^2} max(\tau,\tau^2)$, where $kT_e$ is the plasma temperature, and $\tau$ its optical depth (panel d). This parameter did not evolve significantly before MJD 54759, and then decreased significantly within two days. At the same time, the 20--200 keV flux decreased by $\sim$45\%, and a softening of the spectra occurred.

\subsection{Timing}
The \rxte\ PDS exhibit a shape reminiscent of many other BHs during the first phases of their outbursts. They show a flat top component until a break frequency, above which the decrease in power versus frequency is almost power law like. This is also similar to the PDS of GRS 1915+105 during the so-called $\chi$ class of variability \citep{rodriguez:2008}. We therefore modelled the PDS of \h17\ using the same kind of model: one or two zero-centred Lorentzians, plus narrower ones to account for the presence of QPOs. Due to limited quality of our data, above $\sim$20 Hz the PDS are consistent with white noise. We therefore restricted the fits to the range 0.0156--20 Hz. In all PDS extracted before MJD 54760, a strong type C QPO with its harmonic is visible, as reported by \citet{yamaoka:2008}. The evolution in the QPO parameters (in terms of frequency and RMS amplitude) is reported in Fig. \ref{results} (panels g and h).

The RMS of the continuum remained about $\sim$30\% during the HS (Fig. \ref{results}, panel f). The RMS fraction of the lower frequency QPO remained stable of around 15\%. The RMS fraction of the first harmonic QPO decreased during the first 3 \rxte\ observations, and was insufficiently strong to be measurable afterwards.

\begin{figure}[!ht]
   \centering
   \includegraphics[width=0.39\textwidth]{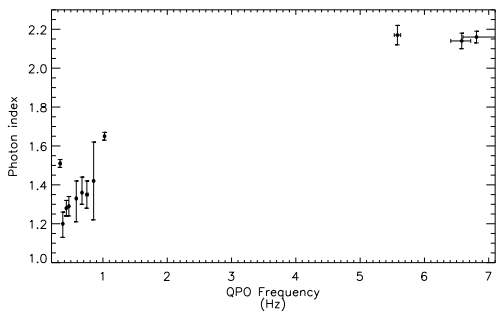}
      \caption{Photon index of the power law component as a function of the frequency of the main QPO. The left part of the diagram corresponds to the HS, while the right part corresponds to the HIMS.}
         \label{correl}
\end{figure}

\section{Discussion}
The spectral and temporal parameters of H1743-322 from MJD 54732 to 54759 indicated that the source was in a HS typical of a Black-Hole Binary \citep{Prat:2008}. This was further supported by the observation of radio emission with a flat or slightly inverted spectrum \citep{Corbel:2008}. We note that the source was observed to be in this state from the very beginning of the outburst. After MJD 54761, the sharp decline in the RMS variability and the sudden increase in the QPO frequency within the timing domain, observed to occur simultaneously with a sudden increase in photon index and decrease in hard X-ray flux, indicate that the source had changed its state. We identify the new state as a HIMS \citep{Belloni:2008, Prat2:2008}.

When comparing this behaviour with that of the beginning of the 2003 outburst, we can see that \h17\ emulated in an approximate way its past behaviour. \h17\ remained within a HS for $\sim$12 days only, compared with $\sim$28 days for the current outburst. The total fluxes were comparable: it was then in the range 2.5--3 $\times$ 10$^{-9}$ \ergcms\ in the 20--100 keV band \citep{McClintock:2007}, compared with 0.5--3.3 $\times$ 10$^{-9}$ \ergcms. The total RMS powers were in the same range, i.e. $\sim$25\% in 2003, compared with $\sim$30\%. The RMS power and photon index are also similar to those exhibited during the 2003 outburst decay, when the source re-entered the HS \citep{Kalemci:2006}. We detect two QPOs, of frequencies $\sim$0.3 and $\sim$1 Hz, and RMS amplitudes of $\sim$5 and $\sim$10\% respectively. During the beginning of the 2003 outburst, a QPO was detected around $\sim$0.1 Hz, along with its first harmonic, of $\sim$3--14\% RMS amplitude.

The main differences reside in the photon indices, which were lower during the 2008 outburst, and the duration of the HS, which was shorter during the 2003 outburst, although it is unclear whether these differences are related.

\medskip
The evolution in the different parameters follows two distinct paths. The overall evolution is characterised by a slowly decreasing hard X-ray flux and a slowly increasing QPO frequency. At the same time, more rapid changes occur: we detect a rapid increase in the flux at the onset of the outburst, and a rapid softening of the spectra precisely as a drastic increase in the QPO frequency takes place at the state transition. These two distinct behaviours are reminiscent of the interpretation of \citet{smith:2002}, which involved two different media: one medium that evolves with a long viscous timescale (the accretion disc), and one that evolves more rapidly (a corona and/or a jet). We can indeed interpret the slowly decreasing hard X-ray flux as being a consequence of the disc evolution: the disc slowly moves in and, as it does, gradually cools the corona. The properties of the QPOs are also comparable to that seen for other BHs in outburst. In particular, the frequency increases with the photon index \citep[Fig. \ref{correl},][]{Vignarca:2003}. The fact that the energy spectra are dominated by the Comptonized component indicates that, although the disc properties probably determine the frequency, the observed QPO power is probably generated in the corona.

The slow increase in the QPO frequency is of particular interest. If we consider a model of an instability propagating inside the disc, the increase in frequency can be interpreted as a movement of the inner part of the disc. In such a model, the frequency is somehow related to the Keplerian rotation frequency, such if the inner part of the disc moves in, the rotation frequency increases, and thus so does the QPO frequency. After the transition to the HIMS, when the QPO frequency increases dramatically, this would indicate that the disk had moved further inwards. Unfortunately, \h17\ was not bright enough to enable the inner radius of the accretion disc to be determined precisely using the multicolor disc blackbody model.

We had the rare privilege to observe a poorly studied early phase of a BH outburst. Our analysis shows that the outburst started in a hard state, initiated by a rapid increase in the flux with a roughly constant spectral shape. The parameters of these early phases could provide important constraint on the subsequent evolution of the outburst. For example, the quite hard spectra seen here could be linked to the length of the HS interval. In the future, more sensitive hard X-ray instruments, such as SIMBOL-X, should allow us to study even fainter and thus earlier phases.

\begin{acknowledgements}
The authors warmly thank the \inte\ planners for having scheduled numerous and regular observations quickly after the beginning of the outburst. We thank P. Varni\`ere for useful discussions on the evolution of the QPOs. MH and JW acknowledge funding from the Bundesministerium f\"ur Wirtschaft und Technologie through Deutsches Zentrum f\"ur Luft- und Raumfahrt grant 50OR0701. This research has made use of data obtained through the High Energy Astrophysics Science Archive Research Center and quick-look results provided by the \rxte/ASM and \swift/BAT teams. Based on observations with \inte, an ESA project with instruments and science data centre funded by ESA member states (especially the PI countries: Denmark, France, Germany, Italy, Switzerland, Spain), Czech Republic and Poland, and with the participation of Russia and the USA. JAT acknowledges partial support from NASA \inte\ Guest Observer grant NNX08AX92G.
\end{acknowledgements}

\bibliographystyle{aa}
\bibliography{H1743}

\end{document}